\def\eiso{E_{\rm iso}}
\def\egamma{E_{\gamma}}
\def\ep{E_{\rm peak}}
\def\epo{E^{\rm obs}_{\rm peak}}
\def\eps{E^{\rm src}_{\rm peak}}

\documentclass[12pt,preprint]{aastex}

\slugcomment{Not to appear in Nonlearned J., 45.}

\shorttitle{X-ray Flash: XRF 050416a}
\shortauthors{Sakamoto et al.}

\begin{document}


\title{Confirmation of the $\eps$ -- $\eiso$ (Amati) relation \\
from the X-ray flash XRF 050416A observed by Swift/BAT}


\author{T. Sakamoto\altaffilmark{1,2}, 
L. Barbier\altaffilmark{1}, 
S. D. Barthelmy\altaffilmark{1}, 
J. R. Cummings\altaffilmark{1,2}, 
E. E. Fenimore\altaffilmark{3},
N. Gehrels\altaffilmark{1}, 
D. Hullinger\altaffilmark{4}, 
H. A. Krimm\altaffilmark{1,5}, 
C. B. Markwardt\altaffilmark{1,4},
D. M. Palmer\altaffilmark{3},
A. M. Parsons\altaffilmark{1},
G. Sato\altaffilmark{6},
J. Tueller\altaffilmark{1},
}

\altaffiltext{1}{NASA Goddard Space Flight Center, Greenbelt, MD 20771}
\altaffiltext{2}{National Research Council, 2101 Constitution Avenue, NW, 
	TJ2114, Washingtion, DC 20418}
\altaffiltext{3}{Los Alamos National Laboratory, P.O. Box 1663, Los
Alamos, NM, 87545}
\altaffiltext{4}{Department of Physics, University of Maryland, 
	College Park, MD 20742}
\altaffiltext{5}{Universities Space Research Association, 10211 Wincopin 
	Circle, Suite 500, Columbia, MD 21044-3432} 
\altaffiltext{6}{Institute of Space and Astronautical Science, 
JAXA, Kanagawa 229-8510, Japan}



\begin{abstract}
We report Swift Burst Alert Telescope (BAT) observations of the X-ray Flash (XRF)
XRF 050416A.  The fluence ratio between the 15-25 keV and 25-50 keV energy bands of 
this event is 1.5, thus making it the softest gamma-ray burst
 (GRB) observed by BAT so far.  
The spectrum is well fitted by the Band function with $\epo$ of 15.0$_{-2.7}^{+2.3}$ 
keV.  Assuming the redshift of the host galaxy (z $=$ 0.6535), the isotropic-
equivalent radiated energy $\eiso$ and the peak energy at the GRB rest frame ($\eps$) 
of XRF 050416A are not only consistent with the correlation found by Amati et al. 
and extended to XRFs by Sakamoto et al., but also fill-in the gap of this relation 
around the 30 -- 80 keV 
range of $\eps$.  This result tightens the validity of the $\eps$ --
 $\eiso$ relation from XRFs to GRBs. 

We also find that the jet break time estimated  
using the empirical relation between $\eps$ and the collimation corrected 
energy $\egamma$ is inconsistent with the afterglow observation 
by Swift X-ray Telescope.  This could be due to the extra external shock 
emission overlaid around the jet break time or to the non existence of a jet break 
feature for XRF, which might be a further challenging for GRB jet emission, 
models and XRF/GRB unification scenarios.  

\end{abstract}





\section{Introduction}

The observations of X-ray flashes (XRF) are providing important 
information for understanding the nature of Gamma-Ray Bursts (GRB).  The 
detailed studies of XRFs started few years ago based on BeppoSAX observations 
\citep{heise2000,kippen2002}, but X-ray rich events had already been 
detected by the $Ginga$ satellite.  \citet{yoshida1989} reported that soft X-ray
emission below 10 keV co-exists with $\gamma$-ray emission 
of GRBs.  About 36\% of the bright bursts observed 
by $Ginga$ have $\epo$ energy, 
which is the photon energy at which the $\nu$F$_{\nu}$ spectrum peaks, around a few keV 
and also show large X-ray to $\gamma$-ray fluence ratios \citep{strohmayer1998}.  

The Wide Field Cameras (WFC) on-board the $Beppo$SAX satellite observed 17 XRFs 
in five years \citep{heise2000}.  
\citet{kippen2002} searched for GRBs and XRFs which were observed in both 
WFC and BATSE.  
The WFC and BATSE joint spectral analysis of XRFs 
shows that their $\epo$ energies are significantly lower than those of the BATSE 
$\epo$ distribution \citep{preece2000}.  
The systematic study of the spectral properties of XRFs observed by $HETE$-2 also 
supports this result \citep{sakamoto2005}.

The afterglow detection and the redshift measurement from the host galaxy of 
XRF 020903, which is one of the softest XRF observed by $HETE$-2, shows the dramatic 
progress in understanding the nature of XRFs.  The prompt emission of XRF 020903 
has $\epo$ $<$ 5.0 keV which is two orders of magnitude smaller than 
that of typical GRBs.  The optical transient and the host galaxy of XRF 020903 
were detected.  Further spectroscopic observation of the host galaxy suggests 
that the redshift is $0.25 \pm 0.01$ \citep{soderberg2004}.  \citet{sakamoto2004} calculated 
the isotropic-equivalent energy $\eiso$ and the peak energy at the source frame 
$\eps$ using the redshift of the host galaxy, and found that XRF 020903 follows 
an extension of the empirical relationship between $\eiso$ and $\eps$ found by 
\citet{amati2002} for GRBs (a.k.a. Amati relation).  
This result provides the observational evidence 
that XRFs and GRBs form a continuum and are a single phenomenon.  

In this paper, we report the prompt emission properties of XRF 050416A observed 
by Burst Alert Telescope (BAT) on-board the $Swift$ satellite.  
The X-ray flash, XRF 050416A, was detected and localized by 
the $Swift$ \citep{gehrels2004} Burst Alert Telescope (BAT;
\citet{barthelmy2005}) at 11:04:44.5
UTC on 2005 April 16 \citep{sakamoto2005b,sakamoto2005c}.  
$Swift$ autonomously slewed to the BAT on-board position, and both $Swift$
X-Ray Telescope (XRT; \citet{burrows2005}) and UV-Optical Telescope 
(UVOT; \citet{roming2005}) detected the afterglow 
(Cusumano et al. (2005) in preparation, Holland et al. (2005) in preparation).  
The afterglow emission of XRF 050416A was also observed by ground 
observatories at various wavelengths \citep{cenko2005a,anderson2005,
li2005,kahharov2005,price2005,cenko2005b,soderberg2005}.   \citet{cenko2005c} 
reported that the host galaxy is faint and blue with large amount of the star 
formation and its redshift is z = 0.6535 $\pm$ 0.0002.  
Throughout this paper, the quoted errors are the 90\% confidence level 
and the sky coordinates in J2000 unless we state otherwise in the text.  

\section{BAT data analysis}

The BAT data analysis was performed using the Swift software package 
(HEAsoft 6.0).  The background was subtracted using the modulations of 
the coded aperture (mask-weighting technique).  In this technique, 
photons with energies higher 
than 150 keV become transparent to the coded mask and these photons 
are treated as a background.  Thus, in this mask-weighted technique 
the effective BAT energy range is from 14 keV to 150 keV.   

\begin{figure}
\centerline{
\includegraphics[scale=0.6,angle=-90]{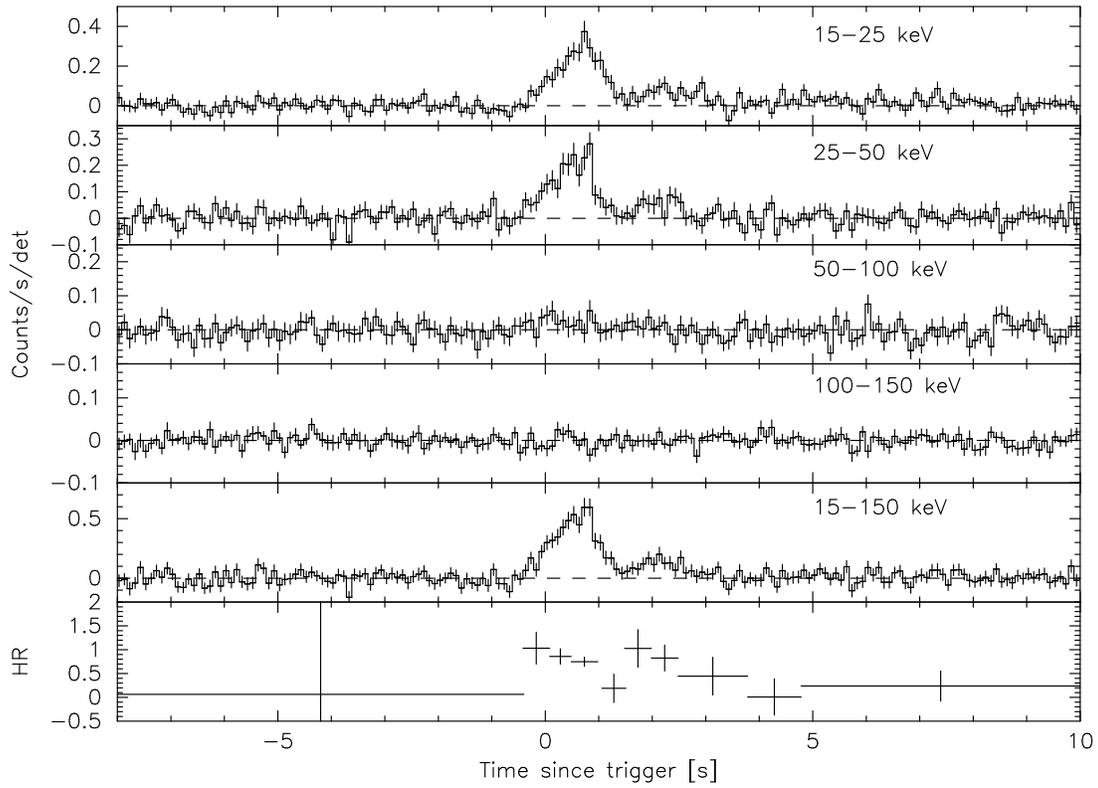}}
\caption{Light curve of XRF 050416a in five energy bands: 15--25 keV, 25--50 keV, 
50--100 keV, 100--150 keV, and 15--150 keV.  The bottom panel shows the hardness 
ratio between the 25--50 keV and 15--25 keV band.}
\label{fig:bat_lc}
\end{figure}

Figure \ref{fig:bat_lc} shows the energy resolved BAT light curves of
XRF 050416A.  It is clear that the signal of the burst is only visible 
below 50 keV.  The burst signal is composed of two peaks.  The 
first peak has a triangular shape with the rise time longer than the
decay time.  When we calculate the spectral lag \citep{norris2000} 
between the 25--50 keV and 15-25 keV band, the cross-correlation 
function lag is $-$0.066$_{-0.018}^{+0.014}$ second (1$\sigma$ error).  
These temporal characteristics are very unusual 
for the typical GRBs (e.g. \citet{mitrofanov1996,norris2000}), 
thus, it is difficult to 
understand them in the frame work of the standard internal shocks models 
in which the rise 
time is always shorter than the decay time and the hard emission 
always exceeds the soft emission (e.g. \citet{piran1999}, \citet{kobayashi1997}).  
The $t_{90}$ and $t_{50}$ in 
the 15-150 keV band are 2.4 and 0.8 seconds, respectively.  This $t_{90}$ 
belongs to the shortest part of the ``long GRB'' classification based 
on the BATSE duration distribution \citep{paciesas1999}.  The fluence 
ratio between the 15-25 keV band and the 25-50 keV band of 1.5 makes 
this burst one of the softest GRBs observed by BAT so far.  The bottom panel of figure
\ref{fig:bat_lc} shows the count ratio between the 25-50 keV and 
15-25 keV bands.  The spectral softening is clearly visible during 
the first and the second peak.  

As reported by the BAT
team\footnote{http://legacy.gsfc.nasa.gov/docs/swift/analysis/bat\_digest.html}, 
we applied the energy-dependent systematic error 
vector in the spectral files before doing any fitting procedure.   
The background subtracted (mask-weighted) spectral data were used in 
the analysis.  The XSPEC v11.3.1 software package was used for 
fitting the data from 14 keV to 150 keV to the model spectrum.  

\begin{figure}
\centerline{
\includegraphics[scale=0.5,angle=-90]{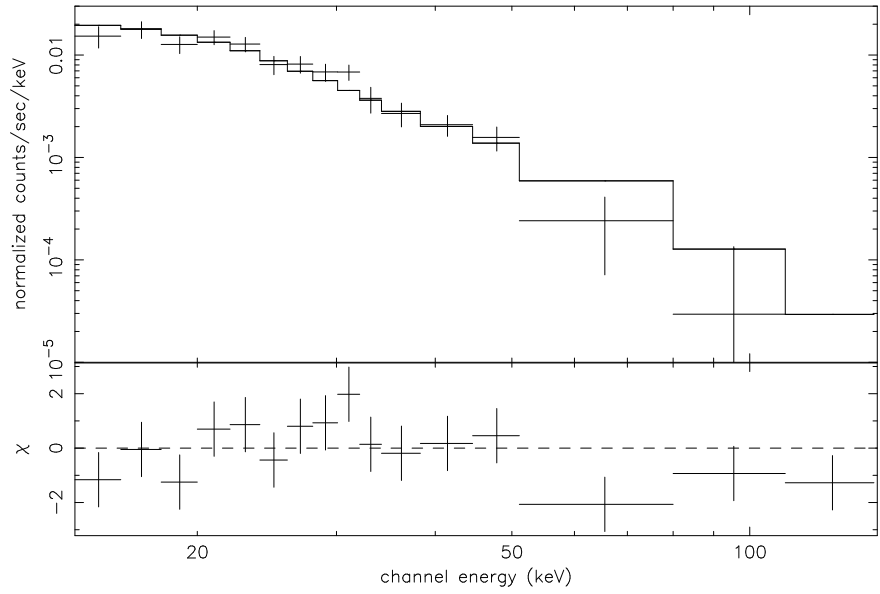}}
\vspace{1cm}
\caption{BAT spectrum of XRF 050416A with a simple power-law model.  
The spectral bins in the figure are binned at least 3 sigma, 
or are grouped in sets of 13 bins.}
\label{fig:bat_spec}
\end{figure}

Table \ref{table:fluence_peakflux} shows the fluences and the peak photon 
fluxes in the various energy bands.  
These fluences and peak photon fluxes were derived directly from fitting 
the time-averaged and 1-s peak spectra respectively assuming the Band 
function with $\alpha = -1$.  
Table \ref{table:spec_para} summarizes the spectral parameters of the BAT 
time-averaged spectrum\footnote{The spectral models which we used throughout 
this paper are following; a simple power-law model (PL): 
$f(E) = K_{30} (E/30)^{\beta}$ and the Band function (Band): 
$f(E) = K_{30} (E/30)^{\alpha} \exp (-E(2+\alpha)/\ep)$,  
if $E < (\alpha - \beta) \ep / (2 + \alpha)$ and 
$f(E) = K_{30}\{(\alpha - \beta)\ep/[30(2+\alpha)]\}^{\alpha-\beta} 
\exp (\beta - \alpha) (E/30)^{\beta}$, if $E \geq (\alpha - \beta) \ep / (2+\alpha)$.}.  
Figure \ref{fig:bat_spec} shows the time-averaged spectrum, accumulated over 
the time interval from $-$0.5 seconds to 3 seconds since the BAT trigger time, 
was fitted with a simple power-law model.  
The photon index $\beta$ which is much steeper 
than $-2$ strongly indicates that the BAT observed the higher energy
part of the Band function 
\citep{band1993}.  Motivated by this result, 
and also by the fact that almost all of GRB and XRF spectra are well 
described by the Band function \citep{preece2000,kippen2002}, 
we tried to fit the spectrum 
with the Band function assuming the low energy photon index
$\alpha$ to be fixed at $-1$, which is the typical value for both GRBs
\citep{preece2000} and XRFs \citep{kippen2002,sakamoto2005}.  The fitting shows a 
significant improvement from a simple power-law model to the Band
function of $\Delta \chi^{2}$ of 7.75 for 1 degree of freedom.  
To quantify the significance of this improvement, we performed 
10,000 spectral simulations assuming our best fit spectral parameters 
in a simple power-law model, and determined in how many cases the 
Band function fit gives 
$\chi^{2}$ improvements of equal or greater than 7.75 over the 
simple power-law.  We found 
equal or higher improvements in $\chi^{2}$ in 62 simulated spectra 
out of 10,000.  Thus, the chance probability of having an equal or higher 
$\Delta\chi^{2}$ of 7.75 with the Band function, when the parent 
distribution is a case of a simple power-law 
model is 0.6\%.  
The observed $E_{\rm peak}$ 
energy, $\epo$, is well constrained at 15.6$_{-2.7}^{+2.3}$  keV, and it 
confirms the soft nature of this burst.  We also applied a $constrained$
Band function fit \citep{sakamoto2004} to the BAT spectrum to estimate 
$\epo$.  The calculated $\epo$ is consistent with the Band function fit of
the fixed $\alpha$ to $-1$: 9.9 keV $<$ $\epo$ $<$ 20.0 keV at the 68\% 
confidence level, 5.1 keV $<$ $\epo$ $<$ 21.8 keV at the 90\% confidence 
level, and $\epo$ $<$ 23.0 keV at the 99\% confidence level.  

\begin{figure}
\centerline{
\includegraphics[scale=0.5,angle=-90]{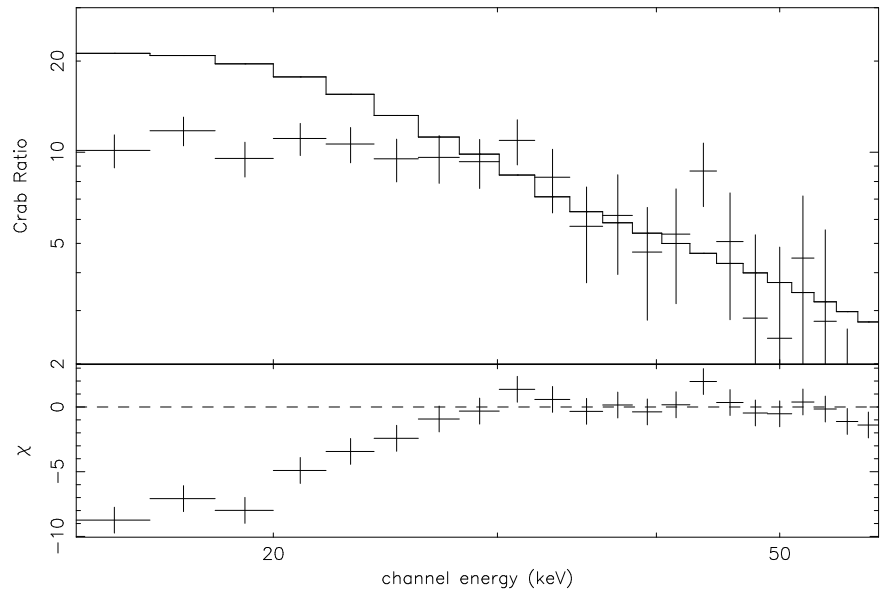}}
\caption{The ratio between the spectral data of XRF 050416A and the Crab nebula.  
The numerator and denumerator of the ratio are the XRF 050416A and the Crab nabula 
spectrum, respectively.  The solid line shows the best fit power-law slope of $-$1.9 
derived from fitting the data above 25 keV.  The bottom panel shows the residuals from this 
best fit power-law slope.  The reduced $\chi^{2}$ is 7.72 in 20 degree
 of freedom.}
\label{fig:crab_ratio}
\end{figure}

The low energy response is crucial for the determination of the
spectral parameters of an XRF and also, as reported by the BAT team
\footnote{Section ``Corrections to Response'' of the BAT Digest 
(http://legacy.gsfc.nasa.gov/docs/swift/analysis/bat\_digest.html)}, 
there is a known problem of $\sim$ 15\% smaller effective area in 
the Crab spectrum below 20 keV when fitting with a pre-launch response 
matrix.  Since the post-launch response matrix which we used in the 
analysis was applying a correction to force the Crab spectrum to 
fit a canonical model 
from 14 keV to 150 
keV, and because we were also also applying the systematic error 
vectors before 
performing the spectral analysis, the systematic effect of this low 
energy problem is very limited.  However, we investigated the
spectrum of XRF 050416A ignoring the spectral bins below 20 keV.  
Even without using spectral bins below 20 keV, the photon index of XRF
050416A is $-3.4 \pm 0.4$, much 
steeper than $-2$ ($\alpha < -2$ at the $>$ 99.99\% confidence level).  
Furthermore, we took the ratio of the spectral data of XRF 050416A 
and the Crab nebula observed at a similar incident angle to XRF 
050416A.  The result is shown in figure \ref{fig:crab_ratio}.  
The flattening trend of the photon index below 25 keV is also clear 
in this figure.  Thus, we conclude that the deviation 
from a simple power-law model below 25 keV is a real 
features of the spectrum of XRF 050416A.  

\section{Discussion}
\begin{figure}
\centerline{
\includegraphics[scale=0.6,angle=-90]{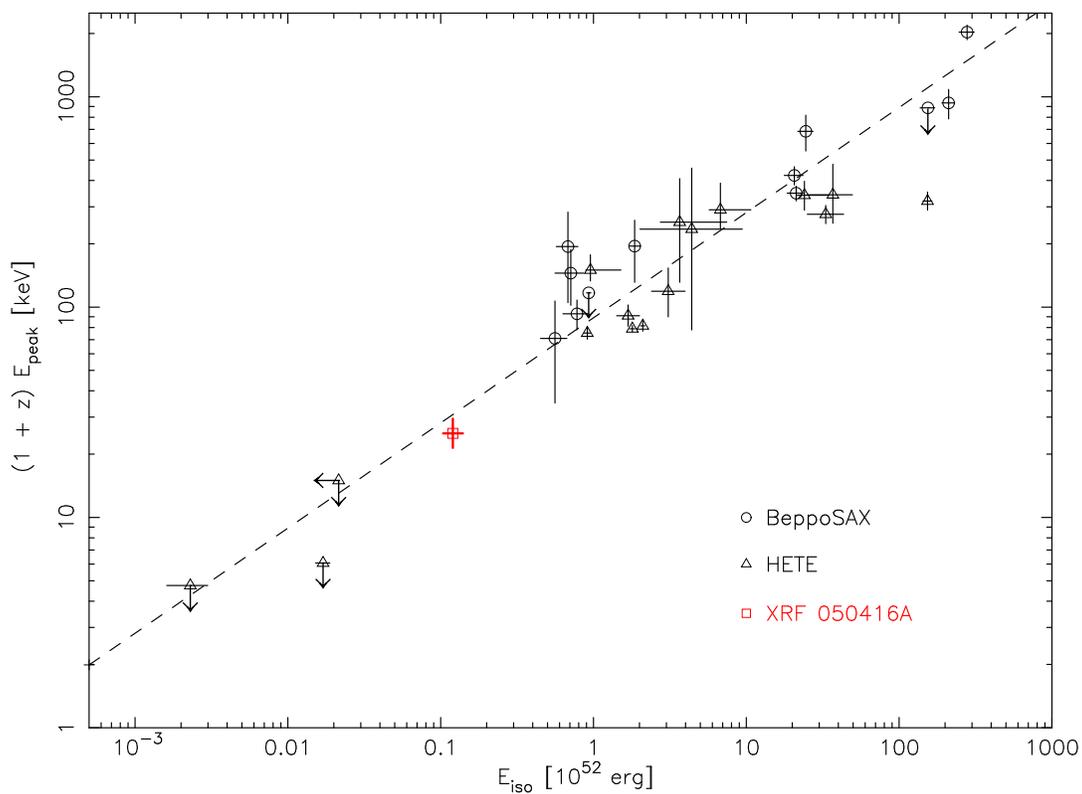}}
\caption{The isotropic-equivalent energy, $\eiso$, versus the peak energy at the GRB 
rest frame, $\eps$, for XRF 050416A (red square) and the known redshift GRBs from $Beppo$SAX 
(circle) and $HETE$-2 (triangle).  The $Beppo$SAX GRB sample is from \citet{amati2002} and the 
$HETE$-2 GRB sample is from Lamb et al.  The dotted line is the 
relation of $\eps$ = 89 ($\eiso/10^{52}\,{\rm erg})^{0.5}$ \citep{amati2002}.}
\label{fig:epeak_eiso}
\end{figure}

%
One of the most important discoveries related to XRF 050416A is the confirmation 
of the $\eps$ -- $\eiso$ relation \citep{amati2002}.  
We calculate the $E_{\rm peak}$ energy at the GRB rest frame, 
$\eps$, and the 
isotropic-equivalent energy (1 -- 10$^{4}$ keV at the rest frame), 
$\eiso$, using the redshift of 
the host galaxy (z=0.6535).  Assuming $\alpha = -1$, 
$\eps$ and $\eiso$ of XRF 050416A are 
25.1$_{-3.7}^{+4.4}$ keV and $(1.2 \pm 0.2) \times 10^{51}$ erg, 
respectively.  
Figure \ref{fig:epeak_eiso} shows the data point of XRF 050416A with 
the known redshift GRBs of $Beppo$SAX and $HETE-2$ sample \citep{amati2003,lamb2004,
sakamoto2004}.  XRF 050416A not only follows the $\eps$
$\propto$ $\eiso^{0.5}$ 
relation, but also fills in the gap of the relation around $\eps$ of 30 -- 80 keV.  
This result tightens 
the validity of this relation at five orders of magnitude in $\eiso$ and at three 
orders of magnitude in $\eps$.  XRF 050416A bridges the gap between XRFs
which have $\eps$ of less than 10 keV and GRBs in the $\eps$ -- $\eiso$ relation.  

The confirmation of $\eps$ -- $\eiso$ relation from XRFs to GRBs gives us 
a clear indication that XRFs and GRBs form a continuum and are a single 
phenomenon.  There are several jet models to explain a unified picture of XRFs and  
GRBs.   The off-axis jet model \citep{yamazaki2004,toma2005}, 
the structured jet model \citep{rossi2002,zhang2002,zhang2004}, 
and the variable jet opening angle model \citep{lamb2005} are the most 
popular models in this aspect.  On the other hand, there are theoretical models to 
explain XRFs in the frame work of the internal shock model \citep{mes2002,mochkovitch2003} 
and of the external shock model \citep{dermer1999,huang2002,dermer2003}.  The 
cited jet models and internal/external shock models not only explain the existences 
of XRFs, under certain assumptions, but also, in some of their realizations 
or for some values of their parameters, they can predict the $\eps$ -- $\eiso$ 
correlation.  

According to the XRT afterglow observation of XRF 050416A, the decay slope 
of the afterglow emission is $\sim$ $-0.9$ from 0.015 days to $\sim$ 34.7 
days after the GRB trigger without any signature of a jet break 
(Cusumano et al. (2005) in preparation; \citet{neusek2005}).  

Using $\eps$ and $\eiso$ of XRF 050416A measured by BAT, we can estimate the 
jet break time using the relation between $\eps$ and the jet collimation-corrected 
energy $E_{\gamma}$ found by \citet{ghirlanda2004} (Ghirlanda relation).  
However, there is a debate about 
the assumption of the jet model used by \citet{ghirlanda2004} to derive 
the relationship between $\eps$ and $E_{\gamma}$ \citep{xu2005,liang2005}.  
Based on this argument, we use the empirical relation between $\eiso$, 
$\eps$, and the jet break time at the rest frame, $t_{\rm jet}^{\rm src}$, 
derived by \citet{liang2005}.  Note that there is no assumption of a jet 
model in the formula found by \citet{liang2005}, and thus their relation is 
purely based on observational properties.  
When we use the equation (5) in \citet{liang2005}, 
($\eiso/10^{52} \,{\rm erg}) = 0.85 \times (\eps/ 100 \,{\rm keV})^{1.94} 
\times (t_{\rm jet}^{\rm src}/1 \, {\rm day})^{-1.24}$, 
the jet break time in 
the observer's frame is estimated to be $\sim$ 1.5 days after the GRB on-set time.  
Note that this estimated jet break time is consistent with the estimation using 
the Ghirlanda relation assuming the circum-burst density of 3 cm$^{-3}$.  
Thus, the estimated jet break time using the empirical $\eps$-$\eiso$-
$t_{\rm jet}^{\rm src}$ relation is 
inconsistent with the null detection of a jet break until more than 34.3 days after 
the trigger by XRT.  

In the off-axis jet model \citep{yamazaki2004,toma2005}, the null detection of 
the jet break in the XRT data of XRF 050416A could be difficult to explain.  
When we assume a 
bulk Lorentz factor of 100, $\eps$ of 300 keV for an on-axis observer, and a jet 
opening angle of 2 degrees, the viewing angle from the jet on-axis is estimated 
to be $\sim$ 4 degrees from the observed $\eps$ of 25 keV.  According to 
\citet{granot2002}, when observing the jet from an angle two times larger 
than the jet opening angle, we would expect to see a rise in the flux 
around one day after the burst.  It is possible to increase the bulk Lorentz factor 
and to reduce the off-axis viewing angle to achieve the same Doppler factor.  However, 
in this case, the afterglow light curve should be close to the on-axis case, thus, 
we would expect to see the jet break around the time we estimated.  

On the other hand, the variable jet opening angle model \citep{lamb2005} might 
work for XRF 050416A if $\egamma$ is a constant value.  If we assume the values 
typical for GRBs ($\eps$ = 
300 keV and the jet opening angle of 5 degrees), the jet opening angle of XRF 
050416A is calculated to be 52 degrees because of the inverse relation between 
$\eps$ and the jet opening angle in the case in which $\egamma$ is a constant.  
When we used the formulation of 
\citet{sari1999} applying the estimated jet opening angle, the 
jet break time will be 64 days in the case of the circum-burst density of 10 
cm$^{-3}$.  Both properties of the low $\eps$ and the null detection 
of the jet break could be explained in the variable jet opening angle model if $\egamma$ 
is constant.  However, as \citet{ghirlanda2004} showed, $\egamma$ 
is not a constant parameter, but has a good correlation with $\eps$.  When we 
applied the Ghirlanda relation, $\eps \propto \egamma^{0.7}$, in the variable 
opening angle model, and re-calculated the jet break time, the break time will 
be 0.7 days assuming the circum-burst density of 10 cm$^{-3}$.  In the variable 
jet opening angle model, there is no way to explain both the Ghirlanda relation 
and the null detection of the jet break by XRT simultaneously.  

One natural way to explain the non-detection of the jet break feature is that 
extra components are overlaid around a jet break time period.  According 
to the afterglow calculations in the X-ray band by \citet{zhang2005}, there are 
several possibilities to hide a jet break feature due to some kinds of emission 
by the external shock.  These are the external shock emission from 
1) the dense clouds surrounding a GRB progenitor (e.g. \citet{lazzati2002}), 
2) a moderately relativistic cocoon component of a two-component jet 
(e.g. \citet{granot2005}) , and 3) a jet with large fluctuations in angular 
direction (patchy jets; \citet{{kumar2000}}).  On the other hand, it might be 
the case that XRFs indeed do not show the signature of a jet break in the afterglow.  
Indeed although the numbers in the sample are limited, there is no clear observational 
indication of a jet break in any XRF afterglow light curve so far.  
If the later case is true, we need to change our view of XRFs completely.  
Thus, the multi-wavelength observations of the XRF afterglows 
will be crucial to investigate whether a jet break feature 
exists in XRFs or not.  

\acknowledgments

We would like to thank R. Yamazaki, and D. Q. Lamb for useful comments 
and discussions.  We would also like to thank the anonymous referee for 
comments and suggestions that materially improved the paper.  
This research was performed while T.S. held a National Research 
Council Research Associateship Award at NASA Goddard Space Flight 
Center.

\clearpage


\begin{table}
\caption{Energy fluences and the peak photon fluxes of XRF 050416A 
assuming the Band function with $\alpha$ = $-1$}
\begin{center}
\begin{tabular}{cccc}\hline
Energy band  & Energy fluence               & Peak photon flux \\
 $[{\rm keV}]$      & [erg cm$^{-2}$]       & [ph cm$^{-2}$ s$^{-1}$] \\\hline
 15 -- 25  & $(1.7 \pm 0.2) \times 10^{-7}$ & $2.9_{-0.3}^{+0.4}$\\ 
 25 -- 50  & $(1.5 \pm 0.2) \times 10^{-7}$ & $1.7 \pm 0.2$\\
 50 -- 100  & $3.4_{-0.6}^{+1.0} \times 10^{-8}$  & $3.2_{-0.4}^{+0.8} \times 10^{-1}$ \\
100 -- 150  & $4.2_{-3.2}^{+11.8} \times 10^{-9}$ & $2.5_{-1.2}^{+3.6} \times 10^{-2}$ \\
 15 -- 150  & $(3.5 \pm 0.3) \times 10^{-7}$      & $5.0 \pm 0.5$ \\\hline
\end{tabular}
\end{center}
\label{table:fluence_peakflux}
\end{table}

\clearpage

\begin{table}
\caption{The time-averaged spectral parameters of XRF 050416A}
\begin{center}
\begin{tabular}{cccccc}\hline
Model & $\alpha$ & $\beta$ & E$_{\rm peak}$ & K$_{30}$ & $\chi^{2}$/d.o.f.\\
      &          &         & [keV]          & [ph cm$^{-2}$ s$^{-1}$
 keV$^{-1}$] &  \\\hline
PL    & & $-3.1 \pm 0.2$ &                  & $(4.3 \pm 0.3)$ $\times 10^{-2}$ & 50.74 / 57\\
PL$^{a}$ & & $-3.4 \pm 0.4$ & & $(4.7 \pm 0.5)$ $\times 10^{-2}$ &
 43.88 / 53\\
Band  & $-$1 (fixed)& $<$ -3.4 & 15.6$_{-2.7}^{+2.3}$ &
 $3.5_{-0.8}^{+1.7}$ $\times 10^{-1}$ & 42.99 / 56\\\hline 
\end{tabular}
\end{center}
\label{table:spec_para}
\tablenotetext{a}{Fitting result using only spectral bins above 20 keV.}
\end{table}


\clearpage

\clearpage

\clearpage

\clearpage

\clearpage

\end{document}